\documentclass[aps,floats,twocolumn]{revtex4}
\usepackage{amssymb}
\usepackage{amsbsy}
\usepackage{epsfig}
\usepackage{amsmath}
\newcommand{\be}{\begin{equation}}
\newcommand{\ee}{\end{equation}}
\newcommand{\bea}{\begin{eqnarray}}
\newcommand{\eea}{\end{eqnarray}}

\newcommand{\Ham}{H}

\begin{document}

\title{Quench dynamics across quantum critical points}
\author{K. Sengupta, Stephen Powell, and Subir Sachdev}
\affiliation{Department of Physics, Yale University, P.O. Box
208120, New Haven, Connecticut 06520-8120}
\date{\today}


\begin{abstract}
We study the quantum dynamics of a number of model systems as
their coupling constants are changed rapidly across a quantum
critical point. The primary motivation is provided by the recent
experiments of Greiner {\em et al.} (Nature {\bf 415}, 39 (2002))
who studied the response of a Mott insulator of ultracold atoms in
an optical lattice to a strong potential gradient. In a previous
work, it had been argued that the resonant response observed at a
critical potential gradient could be understood by proximity to an
Ising quantum critical point describing the onset of density wave
order. Here we obtain numerical results on the evolution of the
density wave order as the potential gradient is scanned across the
quantum critical point. This is supplemented by studies of the
integrable quantum Ising spin chain in a transverse field, where
we obtain exact results for the evolution of the Ising order
correlations under a time-dependent transverse field. We also
study the evolution of transverse superfluid order in the three
dimensional case. In all cases, the order parameter is best
enhanced in the vicinity of the quantum critical point.
\end{abstract}


\maketitle

\section{Introduction}
\label{sec:intro}

Recent experiments with ultracold atoms have achieved reversible
tuning of bosonic atoms between superfluid and Mott insulating
states by varying the strength of periodic potential produced by
standing laser light \cite{Bloch,Kasevich}. The physics of such
ultracold atoms in the Mott insulating state can be described by
bosonic Hubbard model, well known in context of other condensed
matter systems \cite{fwgf,Sachdev1}. However, ultracold atoms in
optical lattices offer much better control over microscopic
parameters of the model. Consequently, it is possible to explore
parameter regimes which are not available in other analogous
condensed matter systems.

This paper will focus on a particular experiment reported by
Greiner {\em et al.} \cite{Bloch}. With the boson system in the
Mott insulating state, they applied a steep potential gradient to
the lattice, and observed its response. In a typical condensed
matter system, one might have expected a response analogous to
that of a sliding charge density wave: no motion of atoms until a
critical tilt was applied, and a sliding motion at all tilts above
the critical tilt. However, the experiment observed strikingly
different behavior: there was a strong {\em resonant} response in
the vicinity of tilts where the potential energy drop between
nearest neighbor optical lattice sites ($E$) equaled the repulsion
between two atoms on the same site ($U$). For $E\sim U$, applying
the tilt produced a noticeable change in the ground state, but (in
contrast to sliding charge density wave systems), there was little
change in the ground state for larger $E$ until a second resonant
peak at $E \sim 2U$. This resonant response is a clear indication
that the atoms experience little extrinsic dissipation, and their
dynamics should be described by an energy-conserving quantum
Hamiltonian.

A framework for describing the experiments of Greiner {\em et al.}
\cite{Bloch} was proposed in Ref.~\onlinecite{Sachdev2} (hereafter
referred to as I). (We also note here the numerical studies of
Braun-Munzinger {\em et al.} \cite{burnett} which addressed these
experiments by studying the time evolution of the underlying
Bose-Hubbard model.) For $w, |E-U| \ll E,U$, where $w$ is the
tunnelling matrix element between nearest neighbor lattice sites,
it was argued that we need only focus on a set of states which
were {\em resonantly coupled} to the original Mott insulating
state. In one dimension, the resonant subspace could be described
simply in terms of nearest neighbor {\em dipole} states,
consisting of a particle and a hole excitation about the Mott
insulator on nearest neighbor states; in higher-dimensions, the
particle and hole were no longer constrained to be on
nearest-neighbor sites but could reside anywhere on planes
orthogonal to the potential gradient, but separated by a single
lattice spacing. An effective Hamiltonian on such resonant
subspaces was proposed in I, and its phase diagram was presented.
In the regime of large potential gradient $E - U > w$, this
effective Hamiltonian possessed ground states with density wave
order with a period of 2 lattice spacings (see also
Ref.~\onlinecite{fendley} for conditions under which other periods
may obtain). It was argued in I that the proximity of the quantum
critical point, associated with the onset of this density wave
order, was responsible for the resonant response observed by
Greiner {\em et al.}.

The tilt experiments of Greiner {\em et al.} were carried out in
highly non-equilibrium situations, and the approach of I was to
describe these, to the extent possible, by an equilibrium analysis
of an effective Hamiltonian describing the primary states accessed
over the experimental time scale. The purpose of the present paper
is to directly address the non-equilibrium dynamics of the tilted
Mott insulator. We will mainly do this using the effective
Hamiltonian of I. The specific question we shall address is the
following. Begin with the system in the ground state in a regime
of small $E=E_i$ where there is no density wave order. Then,
suddenly change the value of $E$ to a $E=E_f$, including values
such that the ground state has density wave order at $E_f$. Allow
the system to evolve under the resulting Hamiltonian. What is the
nature of the state to which the system evolves at long times? We
will find, as conjectured in I, that the density wave order that
develops under this dynamic evolution is most robust when $E_f$ is
near the quantum critical point.

We will also address a similar question for the Ising chain in a
transverse field, $g$. Like the models of I, this model also has a
regime, $g< g_c$, where the ground state has spontaneous Ising
order. However, this much simpler model is completely integrable,
and so offers an opportunity to analyze the non-equilibrium
dynamics exactly. We initialize the Ising model in the ground
state in a transverse field $g_i > g_c$. The transverse field is
then changed rapidly to $g=g_f$, and the wavefunction evolves at
this field. We will compute equal-time correlations in this
wavefunction as a function of the time $t$, including in the $t
\rightarrow \infty$ limit. In some cases, exact closed-form
results will be obtained. The structure of these correlations as a
function of $g_f$ bear some similarity to the results of the model
of I as a function of $E_f$; however, there are some interesting
differences which, we suspect, are related to the integrability of
the Ising chain.

We now outline the remainder of the paper. In Section~\ref{oned}
we present numerical results on  the dynamics of the
one-dimensional dipole model of I. Section~\ref{sec:ising} will
address the non-equilibrium dynamics of the Ising chain in a
transverse field: this analysis uses the Jordan-Wigner
transformation, and obtains the required dynamic correlation
functions in the form of Toeplitz determinants.
Section~\ref{threed} returns to the model of I, but turns to the
dynamics in three dimensions; here, we use a combination of
mean-field theory and exact diagonalization to obtain results
similar to those in Section~\ref{oned}, but with the order
parameter now being a `transverse superfluid' order. We review the
results and discuss implications for experiments in
Section~\ref{conc}.

\section{Dipole dynamics in one dimension}
\label{oned}

This section will describe our numerical results on the quantum
dynamics of the one-dimensional dipole model of the Mott insulator
in a potential gradient.

Starting from a parent Mott state with $n_0$ bosons per site, we
identified the set of states which are resonantly coupled to the
parent Mott state when $U \sim E$ (recall that $U$ is the
repulsive energy between two bosons on the same site, and $E$, the
`electric field', is the potential drop between two nearest
neighbor sites). In one dimension, the resonant subspace involves
dipole states consisting of quasihole-quasiparticle pairs at
adjacent sites, and the low energy behavior of the system can be
described by the effective dipole Hamiltonian obtained in I:
\begin{eqnarray}
H_{\rm 1D}[E] &=& -w \sqrt{n_0(n_0+1)}\sum_\ell (d_\ell^\dagger +
d_\ell)
\nonumber\\
&& +(U-E) \sum_{\ell} d_\ell^{\dagger} d_\ell . \label{ham1d}
\end{eqnarray}
The dipoles are subject to hardcore constraints that there is
never more than a single dipole on any pair of nearest neighbor
sites
\begin{equation}
d_\ell^{\dagger} d_\ell \le
1~~~;~~~d_{\ell+1}^{\dagger}d_{\ell+1}d_\ell^{\dagger} d_\ell = 0.
\end{equation}
When the electric field $E$ is adiabatically tuned through $U$,
the ground state of the system changes from one with no dipoles
($U\gg E$) to one with maximum possible number of dipoles ($E\gg
U$). At an intermediate critical electric field
\begin{equation}
E_c = U + 1.310 w \sqrt{n_0 (n_0 + 1)},
\end{equation}
the system undergoes a quantum phase transition in the Ising
universality class.

As discussed in Section~\ref{sec:intro}, we study the dynamics of
the ultracold atoms when the potential gradient is changed
suddenly. Such a situation can be very easily achieved
experimentally in these systems by rapidly shifting the center of
the confining magnetic trap. We shall specifically consider the
situation where the change in the potential gradient is fast
enough for the sudden perturbation assumption to be valid but slow
enough to restrict the dynamics within the resonant subspaces so
that the Hamiltonians (\ref{ham1d}) (and (\ref{ham3d}) in
Section~\ref{threed}) are still valid.

We assume that the atoms in the 1D lattice are initially in the
ground state $|\Psi_G\rangle $ of the dipole Hamiltonian
(\ref{ham1d}) with $E=E_{i} \ll E_c$. This ground state
corresponds to dipole vacuum. Consider shifting the center of the
magnetic trap so that the new potential gradient is $E_{f}$. If
this change is done suddenly, the system initially remains in the
old ground state. The state of the system at time $t$ is therefore
given by
\begin{eqnarray}
|\Psi(t)\rangle  =\sum_n c_n \exp(-i\epsilon_n t/\hbar) |n\rangle
, \label{state}
\end{eqnarray}
where $|n\rangle $ denotes the complete set of energy eigenstates
of the Hamiltonian $H_{\rm 1D}[E_{f}]$ in (\ref{ham1d}),
$\epsilon_n = \,\,\langle n|H_{\rm 1D}[E_{f}]|n\rangle $ is the
energy eigenvalue corresponding to state $|n\rangle $, and $c_n
=\,\,\langle n|\Psi(t=0)\rangle  = \langle n|\Psi_G\rangle $
denotes the overlap of the old ground state with the state
$|n\rangle $. Notice that the state $|\Psi(t)\rangle $ is no
longer the ground state of the new Hamiltonian. Furthermore, in
the absence of any dissipative mechanism, which is the case for
ultracold atoms in optical lattices, $|\Psi(t)\rangle $ will never
reach the ground state of the new Hamiltonian. Rather, in general,
we expect the system to thermalize at long enough times, so that
the correlations are similar to those of $H_{\rm 1D} [E_f]$ at
some finite temperature.

We are now in a position to study the dynamics of the Ising
density wave order parameter
\begin{equation}
O = \frac{1}{N}\langle \Psi|\sum_\ell (-1)^\ell d_\ell^{\dagger}
d_\ell|\Psi\rangle, \label{isingO}
\end{equation}
where $N$ is the number of sites. The time evolution of $O$ is
given by
\begin{eqnarray}
O(t) &=& \frac{1}{N} \sum_{m,n} c_m  c_n \cos\left[\left( E_m-E_n
\right)t/\hbar \right] \nonumber\\
&& \times  \langle m|\sum_\ell (-1)^\ell d_\ell^{\dagger}
d_\ell|n\rangle \label{ising}
\end{eqnarray}
Eq.\ \ref{ising} is solved numerically using exact diagonalization
to obtain the eigenstates and eigenvalues of the Hamiltonian
$H_{\rm 1D}[E_{f}]$. Before resorting to numerics, it is however
useful to discuss the behavior of $O(t)$ qualitatively. We note
that if $E_f$ is close to $E_i$, the old ground state will have a
large overlap with new one $\it i.e.$ $c_m \sim \delta_{m1}$.
Hence in this case we expect $O(t)$ to have small oscillations
about $O(t=0)$. On the other hand, if $E_f \gg E_c$, the two
ground states will have very little overlap, and we again expect
$O(t)$ to have a small oscillation amplitude. This situation is in
stark contrast with the adiabatic turning on of the potential
gradient, where the systems always remain in the ground state of
the new Hamiltonian $H_{\rm 1D}[E_f]$, and therefore has a maximal
value of $\langle O\rangle $ for $E_f \gg E_c$. In between, for
$E_f \sim E_c$, the ground state $|\Psi\rangle $ has a finite
overlap with many states $|m\rangle $, and hence we expect $O(t)$
to display significant oscillations. Furthermore, if the symmetry
between the two Ising ordered states is broken slightly (as is the
case in our studies below), the time average value of $O(t)$ will
be non-zero.

\begin{figure}
        \centerline{\epsfig{file=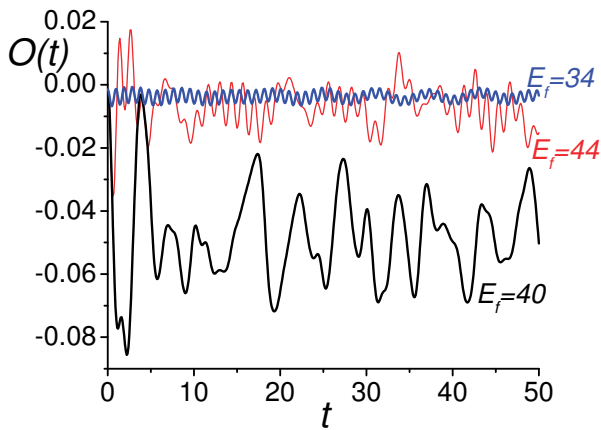,width=8cm,angle=0}}
        \caption{Evolution of the Ising order parameter in (\protect\ref{isingO})
        under the Hamiltonian $H_{\rm 1D} [E_f]$ for $n_0 = 1$.
        The initial state is the ground state of $H_{\rm 1D} [E_i
        ]$. All the plots in this section have
        $U=40$, $w=1$, and $E_i=32$, and consequently the
        equilibrium quantum critical point is at $E_c = 41.85$.}
\label{fig1}
\end{figure}

This qualitative discussion is supported by numerical calculations
on finite size systems for system size $N=9,11,13$. For numerical
computations with finite systems, we choose systems with an odd
number of sites and open boundary conditions, so that dipole
formation on odd sites is favored, thus breaking the $Z_2$
symmetry. The results are shown in Figs.\ \ref{fig1}-\
\ref{fig3a}.
\begin{figure}
        \centerline{\epsfig{file=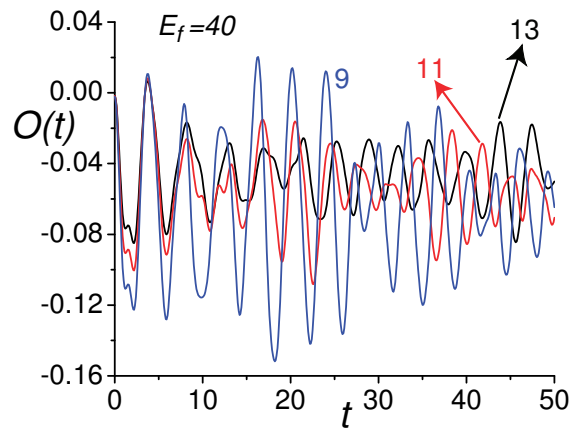,width=8cm,angle=0}}
        \caption{System size ($N$) dependence of the results of Fig~\protect\ref{fig1} for
        $E_f=40$. The curves are labelled by the value of $N$. }
         \label{fig2}
\end{figure}
\begin{figure}
        \centerline{\epsfig{file=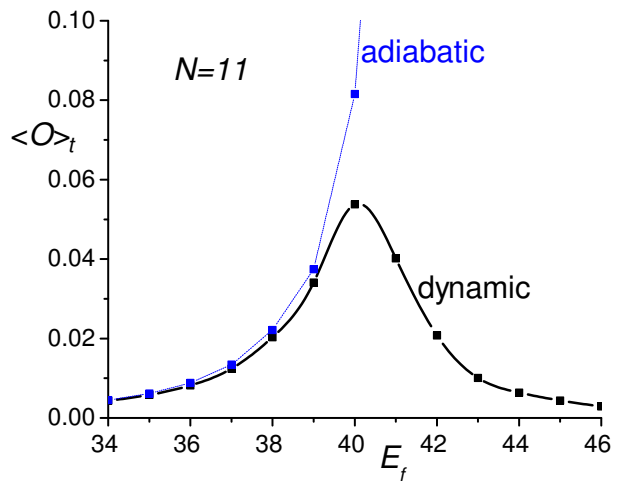,width=8cm,angle=0}}
        \caption{The curve labelled `dynamic' is the long time
        limit, $\langle O\rangle_t$
        of the Ising order in (\protect\ref{ising}) as a function
        of $E_f$ (for $N=11$), with other parameters as in
        Fig~\protect\ref{fig1}. This long time limit can be obtained
        simply by setting $m=n$ in (\protect\ref{ising}).
        For comparison, in the curve
        labelled `adiabatic', we show the expectation value of the
        Ising order $O$ in the ground state of $H_{\rm 1D} [E_f]$;
        such an order would be observed if the value of $E$ was
        changed adiabatically. Note that the dynamic curve has its
        maximal value near (but not exactly at) the equilibrium quantum critical point
        $E_c = 41.85$, where the system is able to respond most easily to
        the change in value of $E$; this dynamic curve is our theory of
        the `resonant' response in the experiments of Ref.~\protect\onlinecite{Bloch}
        discussed in Section~\ref{sec:intro}. In contrast the adiabatic
        result {\em increases monotonically} with $E_f$ into the $E>E_c$ phase where
        the Ising symmetry is spontaneously broken.}
        \label{fig3}
\end{figure}
\begin{figure}
        \centerline{\epsfig{file=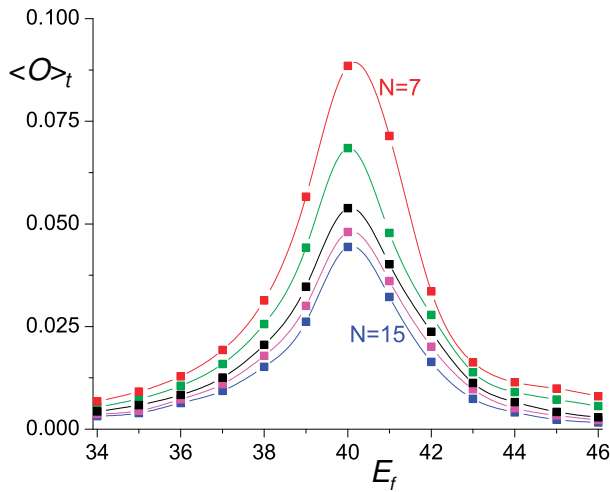,width=8cm,angle=0}}
        \caption{Size dependence of the `dynamic' results in
        Fig~\protect\ref{fig3}. The sizes range from $N=7$ to $N=15$ (as labeled),
        with the intermediate values $N=9,11,13$:
        $\langle O\rangle_t$ decreases monotonically with $N$}
        \label{fig3a}
\end{figure}
Fig.\ \ref{fig1} shows the oscillation of the order parameter
$O(t)$ for different values of $E_f$ for $N=13$. In agreement with
our qualitative expectations, the oscillations have maximum
amplitude when $E_f \approx 40$ is near the critical value $E_c =
41.85$. For either $E_f \ll E_c$ or $E_f \gg E_c$, the
oscillations have a small amplitude around $O(t=0)$. Furthermore,
it is only for $E_f \approx E_c$ that the time-average value of
$O(t)$ is appreciable. Fig.\ \ref{fig2} shows the system size
dependence of the time evolution for $E_f = U=40$. We find that
the oscillations remain visible as we go to higher system sizes,
although they do weaken somewhat. More significantly, the time
average value of $O(t)$ remains non-zero, and has a weaker
decrease with system size. In Fig.\ \ref{fig3}, we plot the long
time limit of the Ising order parameter, $\langle O\rangle _t$, as
a function of $E_f$, and compare it with the $O_{\rm ad}$, the
value of the order parameter when $E$ reaches $E_f$ adiabatically
and the wavefunction is that of the ground state at $E=E_f$. We
find that $\langle O\rangle _t$ stays close to $O_{\rm ad}$ as
long as there is a large overlap with between the old and the new
ground states. However, as we approach the adiabatic phase
transition point, this overlap decreases and $\langle O\rangle _t$
can not follow $O_{\rm ad}$ any more. The deviation of $\langle
O\rangle _t$ is therefore a signature that the system is now in a
different phase for the new value of the electric field.

The `dynamic' curve in Fig~\ref{fig3} should be compared with Figs
5e,f in Ref.~\onlinecite{Bloch}. The latter show that the Mott
insulator has a resonantly strong response to an applied potential
gradient $E \sim U$. Here, we have found a similar resonant
enhancement in a simple model system in one dimension, induced by
the proximity of a quantum critical point.

We comment briefly on the nature of the thermodynamic limit, $N
\rightarrow \infty$ for the results in Fig~\ref{fig2},~\ref{fig3}.
For $O_{\rm ad}$ it is clear that there is a non-zero limit only
for $E > E_c$, when it equals the order parameter of the
spontaneously broken Ising symmetry. If we assume that the system
thermalizes at long times for the dynamic case, then $\langle O
\rangle_t$ corresponds to the expectation value of the equilibrium
order parameter in $H_{1D} [E_f]$ at some finite temperature. In
one dimension, it is not possible to break a discrete symmetry at
finite temperatures, and so the thermodynamic limit of the order
parameter must always vanish. By this reasoning, we expect
$\langle O \rangle_t$ to also vanish in the thermodynamic limit.
This is consistent with the results in Fig.\ \ref{fig3a}, where we
show the $N$ dependence of the long time limit $\langle O
\rangle_t$. Our data are at present not extensive enough to
definitely characterize the dependence of $\langle O \rangle_t$ on
$N$.

\section{Dynamics of the quantum Ising chain}
\label{sec:ising}

As a complement to the physically relevant, but numerical,
computations in Section~\ref{oned}, this section will describe
similar results in a simpler, analytically tractable model. We
will consider the integrable Ising chain in a transverse field,
which also has a zero temperature, quantum phase transition
between a phase with a broken $Z_2$ symmetry and a symmetric
phase. We will address questions on the evolution of the
wavefunction under a time-dependent change in the transverse
field.

The model of interest in this section is
\begin{equation}
H_I = -J \sum_{j} \left( \sigma^z_j \sigma^z_{j+1} + g(t)
\sigma^x_i \right), \label{HI}
\end{equation}
where $\sigma^{x,z}_j$ are Pauli matrices acting on a `spin' on
the sites, $j$, of an infinite chain. We have allowed the
transverse field to acquire an arbitrary time dependence $g(t)$.
We will mainly consider here the case of a sudden change at time
$t=0$ from an initial value $g(0^-) = g_i$ to a final value
$g(0^+)=g_f$, but our methods easily generalize to the arbitrary
time dependence in $g(t)$.

For time-independent $g(t)$, $H_I$ has a quantum critical point at
$g=g_c = 1$, with two equivalent ground states for $g< g_c$
related by a global $Z_2$ spin-flip. However, unlike
Section~\ref{oned} we will not introduce any external perturbation
which introduces a preference between these two states: all such
perturbations destroy the integrability of $H_I$. Consequently, we
do not obtain any useful information from the analog of the
time-dependence of the order parameter in (\ref{isingO}),
(\ref{ising}), as these quantities will be identically zero at all
times. Rather, we will compute here the two-point correlation
function of the order parameter in an infinite chain, which is
\begin{equation}
G_n (t) = \langle \psi (t) | \sigma^z_j \sigma^z_{j+n} | \psi (t)
\rangle. \label{defgn}
\end{equation}
Here $|\psi (t) \rangle$ is the state of the system at time $t$,
evolving under the Sch\"odinger equation specified by the
time-dependent Hamiltonian $H_I$. In equilibrium, the information
contained in a correlation function like (\ref{defgn}) is related
to an observable like that in (\ref{ising}) (which is the response
in the Ising order parameter to perturbations in the boundary
condition) by the fluctuation-dissipation theorem. However, we are
not aware of any analog of such a theorem for the non-equilibrium
case under consideration here, and so are not able to directly
relate the results of the present section to those of
Section~\ref{oned}.

Our analysis of $H_I$ proceeds with the standard Jordan-Wigner
transformation, and we follow the notation and methods of Chapter
4 of Ref.~\onlinecite{Sachdev1}. We express the $S=1/2$ states in
terms of those of the spinless Jordan-Wigner fermion $c_j$, and
after transforming to momentum space fermions $c_k$, the
Hamiltonian becomes
\begin{eqnarray}
H_I &=& J \sum_k\left[ 2\left(g-\cos k \right) c_k^{\dagger} c_k
\right. \nonumber \\ &~&~~~~~~~~~~\left. - i \sin k
\left(c^{\dagger}_{-k} c_{k}^{\dagger} + c_{-k} c_{k} \right) -g
\right].
\end{eqnarray}
Now, transforming to the Heisenberg picture, we can follow the
evolution of the system by solving the equations of motion
\begin{equation}
\frac{d c_k}{dt} = i \left[ H_I , c_k \right].
\end{equation}
These equations are easily solved by a Bogoliubov transformation.
Finally, the correlator in (\ref{defgn}) is computed by a simple
generalization of the methods appropriate for the equilibrium
case. A few details of such a computation appear in the Appendix.

Here, we discuss the results for $G_n (t)$ for the case of a
sudden change from $g (0^-) = g_i$ to $g(0^+ ) = g_f$. For $t< 0$,
we assume the system is in the ground state appropriate for
$g=g_i$, and consequently $G_n (t<0)$ is independent of $t$ and
equal to the well-known equilibrium result at $g=g_i$. For $t>0$,
there is a non-trivial time dependence, and it is possible to
obtain the general expression for $G_n (t)$ as described in the
Appendix. We will restrict our attention here to the simpler
expression of the long time limit $G_n (t \rightarrow \infty)$,
which is the primary quantity of physical interest. For this, we
obtain the Toeplitz determinant
\begin{equation} G_n (\infty) =
\begin{vmatrix}
a_0 & a_{-1} & \ldots & a_{-n+1}\\
a_1 & a_0 & \ldots & a_{-n+2}\\
\vdots & \vdots & \ddots & \vdots\\
a_{n-1} & a_{n-2} & \ldots & a_0
\end{vmatrix},
\label{toep}
\end{equation}
where
\begin{equation}
a_r = \frac{1}{2\pi}\int_{-\pi}^{\pi} e^{-ikr} \tilde{a}(k),
\end{equation}
with
\begin{eqnarray}
\tilde{a}(k) &=& \frac{2(g_f g_i + 1)z - (g_f+g_i)(z^2+1)}{2(z-g_f)} \nonumber \\
&~&~~~~~~~~~~~\times \sqrt{\frac{z}{(z-g_i)(1-z g_i)}},
\label{tak}
\end{eqnarray}
where $z = e^{ik}$.

We now need to evaluate the $n \times n$ Toeplitz determinant in
(\ref{toep}), especially for the case of large $n$. In the
equilibrium situation, this is aided by Szeg\"o's lemma, and its
generalization in the Fisher-Hartwig formula \cite{fh}. For the
present situation, the expression in (\ref{tak}) does not obey the
winding number constraint required for application of the
Fisher-Hartwig formula, and so we are unable to take advantage of
this result. However, we shall show that an exact evaluation of
(\ref{toep}) is possible for two important special cases ($g_i =
\infty$ and $g_i = 0$), and supplement these by numerical
evaluation of (\ref{toep}) for other values of $g_i$.

In the case $g_i = 0$, we have
\begin{equation}
\tilde{a}(k) = \frac{2z - g_f (z^2+1)}{2(z-g_f)}
\end{equation}
and it is straightforward to evaluate $a_r$ by contour
integration. This gives
\begin{equation}
\begin{array}{c|c|c}
 & g_f < 1 & g_f > 1\\ \hline \rule[-4mm]{0mm}{11mm}
r \le -1 &  \displaystyle \frac{g_f^{-r}}{2} (1-g_f^2) & 0\\
\hline \rule[-4mm]{0mm}{11mm} r = 0 & 1 -  \displaystyle
\frac{g_f^2}{2} &  \displaystyle \frac{1}{2}\\ \hline
\rule[-4mm]{0mm}{11mm} r = 1 & -
\displaystyle \frac{g_f}{2} & - \displaystyle \frac{1}{2g_f}\\
\hline \rule[-4mm]{0mm}{11mm} r \ge 2 & 0 &  \displaystyle
\frac{g_f^{-r}}{2} (g_f^2 - 1)
\end{array}~~.
\end{equation}

For the case $g_i = +\infty$, we have
\begin{equation}
\tilde{a}(k) = \frac{2g_f z - (z^2+1)}{2(z-g_f)}
\end{equation}
and $a_r$ is given by
\begin{equation}
\begin{array}{c|c|c}
 & g_f < 1 & g_f > 1\\ \hline \rule[-4mm]{0mm}{11mm}
r \le -1 &  \displaystyle -\frac{g_f^{-r-1}}{2} (1-g_f^2) & 0\\
\hline \rule[-5mm]{0mm}{11mm} r = 0 &  \displaystyle \frac{g_f}{2}
&  \displaystyle \frac{1}{2g_f}\\ \hline \rule[-5mm]{0mm}{11mm} r
= 1 &  \displaystyle -\frac{1}{2} & \displaystyle \frac{1}{2g_f^2}
- 1\\ \hline \rule[-5mm]{0mm}{11mm} r \ge 2 & 0 & \displaystyle
-\frac{1}{2g_f^{r+1}}(g_f^2 - 1)
\end{array}~~.
\end{equation}

\newtheorem{cond}{Condition}
In both of these two cases, the following conditions are met:
\begin{cond}
For $g_f > 1$, $a_r = 0$ for $r \le -1$.
\end{cond}
\begin{cond}
For $g_f < 1$, $a_r = 0$ for $r \ge 2$.
\end{cond}
\begin{cond}
For $g_f < 1$, $a_r = ga_{r+1}$ for $r \le -2$.
\end{cond}

Using Condition 1, we can immediately write $G_n (\infty ) =
a_0^n$ for $g_f
> 1$. For $g_f < 1$, define \be D_n^r =
\begin{vmatrix}
a_{-r} & a_{-r-1} & \ldots & a_{-r-n+1}\\
a_{1} & a_0 & \ldots & a_{-n+2}\\
\vdots & \vdots & \ddots & \vdots\\
a_{n-1} & a_{n-2} & \ldots & a_0
\end{vmatrix},
\ee so that $G_n (\infty ) = D_n^0$. Condition 2 gives $D_n^r =
a_-r D_{n-1}^0 - a_1 D_{n-1}^{r+1}$ and Condition 3 gives $D_n^r =
g_f D_n^{r-1}$ for $r \ge 2$. Also, $D_1^r = a_{-r}$. We can
therefore write \bea
\begin{pmatrix}
D_n^0\\
D_n^1
\end{pmatrix}
&=&
\begin{pmatrix}
a_0 & -a_1\\
a_{-1} & -a_1 g_f
\end{pmatrix}
\begin{pmatrix}
D_{n-1}^0\\
D_{n-1}^1
\end{pmatrix}\\
&=& {\begin{pmatrix}
a_0 & -a_1\\
a_{-1} & -a_1 g_f
\end{pmatrix}}^{n-1}
\begin{pmatrix}
D_1^0\\
D_1^1
\end{pmatrix}\\
&=& {\begin{pmatrix}
a_0 & -a_1\\
a_{-1} & -a_1 g_f
\end{pmatrix}}^n
\begin{pmatrix}
1\\
0
\end{pmatrix}.
\eea This can be evaluated by diagonalizing the matrix.

Collecting all the analytic results above, we have for the case
$g_i = 0$:
\begin{equation}
G_n (\infty )= \left\{
\begin{array}{r} \displaystyle
\frac{g_f^{n+1}}{2^n} \cosh \left[ (n+1) \ln \left(
\frac{1+\sqrt{1-g_f^2}}{g_f} \right) \right], \\
\mbox{for $g_f \leq 1$} \\~\\ \displaystyle
\left(\frac{1}{2}\right)^n,~~~~~~~~~~~\mbox{for $g_f \geq 1$}
\end{array}
\right.
\end{equation}
In the limit that $n \rightarrow \infty$, the result for $g_f < 1$
becomes \be G_n (\infty ) \rightarrow {\left(
\frac{1+\sqrt{1-g_f^2}}{2} \right)}^{n+1}. \ee

In the case $g_i = +\infty$, the corresponding results are
\begin{equation}
G_n (\infty )= \left\{
\begin{array}{cc} \displaystyle
\left(\frac{1}{2}\right)^n \cos \left( n \arccos (g_f) \right), &
\mbox{for $g_f \leq 1$} \\~\\ \displaystyle
\left(\frac{1}{2g_f}\right)^n, & \mbox{for $g_f \geq 1$}
\end{array}
\right. \label{e1}
\end{equation}
Note that there are spatial oscillations in the correlator for the
case where the field is reduced from a large positive value ($g_i
= +\infty$) to a value below the critical point ($g_f<1$).

Of these exact results, the case $g_i = \infty$ is the one that
corresponds most closely to the physical situation discussed in
Section~\ref{oned}. Here we start from an fully `disordered'
initial state, and then suddenly change parameters to values with
increasing order (this is the analog of increasing $E$ in
Section~\ref{oned}). For final parameter values $g_f > g_c = 1$,
we find here a result quite similar to that found in
Section~\ref{oned}: from (\ref{e1}) we see that the order
parameter correlations decay with the a correlation length $\xi_f$
given by
\begin{equation}
\xi_f = \frac{1}{\ln(2 g_f )}.
\end{equation}
This increases monotonically with decreasing $g_f$, and is thus
similar to the increase in the value of $\langle O \rangle_t$ with
increasing $E_f$ for $E_f < E_c$ in Fig~\ref{fig3}. By the analogy
with Fig~\ref{fig3}, we would expect here that there is a maximum
in $\xi_f$ at $g=g_c$. However, we find a somewhat different
behavior for $g_f < g_c$ in (\ref{e1}): the correlations do not
decay in a simple exponential, but now oscillate, with the period
of oscillation becoming smaller with decreasing $g_f$. So the
correlations of the Ising ordered state are indeed best formed at
$g_f = g_c$, but we find an unusual oscillatory decay of
correlations for $g_f < g_c$. The oscillations are a clear
indication of the absence of thermalization in the present model,
and we expect they are special consequence of its integrability.

\begin{figure}
        \centerline{\epsfig{file=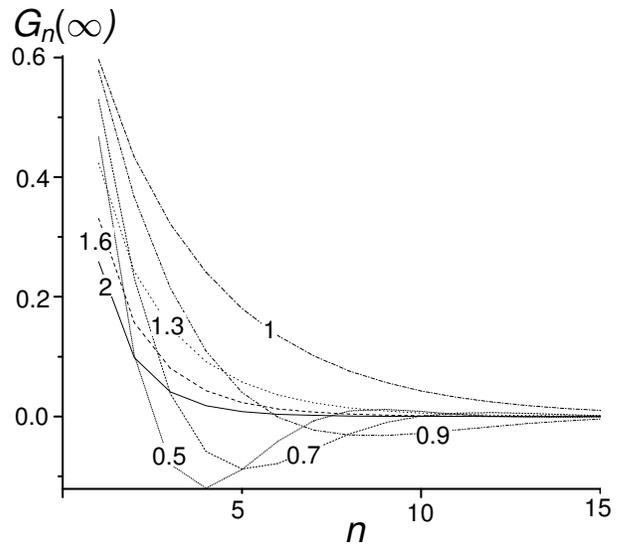,width=8cm,angle=0}}
        \caption{Ising order correlations defined in (\protect\ref{defgn}). The
        system is in the ground state of $H_I$ for $t < 0$ with $g=g_i = 2$.
        At $t=0^+$, the value of $g$ is changed suddenly to $g=g_f$, and remains
        at this value for all $t > 0$. Note that at long times, the order is best
        developed for $g_f = 1$, which is the location of the equilibrium quantum
        critical point. This result is the analog of Figs~\protect\ref{fig3} and~\ref{fig3a}
        for the dipole model of Section~\protect\ref{oned}.}
        \label{fig3b}
\end{figure}

We extended these analytic results by numerical evaluation of
(\ref{toep}) for other values of $g_i$, and found closely related
behavior. Our results for $g_i = 2$ are shown in Fig~\ref{fig3b},
and these are the analog here of the results in Fig~\ref{fig3}
and~\ref{fig3a}.
As $g_f$ is decreased, the correlations become longer-ranged,
until they reach a maximum range at $g_f = g_c = 1$. At smaller
values of $g_f$, the correlations acquire an oscillatory behavior,
but are also clearly shorter ranged. So the Ising order is best
developed for $g_f$ near the quantum critical point.

\section{Dynamics in three dimensions}
\label{threed}

We now return to the `tilted' Mott insulator problem addressed in
Section~\ref{oned} and in I. Here we will address questions of
quench dynamics for the three dimensional case. As discussed at
length in I, the resonant subspace in 3D is described by
quasiparticles and quasiholes which are free to move in the
directions transverse to the applied electric field. Consequently,
the dipoles of Section~\ref{oned}, which are bound
quasihole-quasiparticle pairs in adjacent sites, constitute only a
small part of the resonant subspace, and an effective Hamiltonian
for unbound quasiparticle and quasihole states is necessary. A
mean-field theory of this effective Hamiltonian was examined in I,
and a fairly complex phase diagram was found. In addition to the
Ising density wave order that appeared in one dimension, states
with a {\em transverse superfluid} order were present. The latter
states correspond to delocalization of the quasiholes and
quasiparticles in the direction transverse to the applied electric
field.

In this section, we will address the quench dynamics across the
transition associated with the onset of transverse superfluid
order. This was found to be a second-order transition in the
mean-field theory of I, and here we will extend the mean-field
theory to an analysis of the non-equilibrium dynamics across the
superfluid-insulator transition. We will not examine here the
onset of Ising order, already studied in Section~\ref{oned}; the
present mean-field theory found a strong first-order transition
for the onset of Ising order. Our analysis will be restricted to
the regime where both the superfluid and insulating states have no
Ising density wave order.

The effective mean-field Hamiltonian describing the dynamics of
these quasiparticles and quasiholes can be written as in I:
\begin{eqnarray}
\lefteqn{ H_{3D}[\langle p_\ell\rangle ,\langle h_\ell\rangle ;E]=}  \nonumber\\
&& \sum_\ell \Bigg [ -w Z \Big(n_0 h_\ell \langle h_\ell\rangle ^*
+ (n_0+1)p_\ell \langle p_\ell\rangle ^*
\nonumber\\
&&+ {\rm h.c} \Big) -w \sqrt{n_0(n_0+1)} \Big( p_\ell h_{\ell-1}
+{\rm h.c} \Big)
\nonumber\\
&& + \frac{(U-E)}{2} \left(p_\ell^{\dagger} p_\ell +
h_\ell^{\dagger} h_\ell \right)  -\mu_\ell
\left(p_{\ell+1}^{\dagger} p_{\ell+1} - h_\ell^{\dagger}
h_\ell \right) \Bigg]. \nonumber\\
\label{ham3d}
\end{eqnarray}
Here $\ell$ is a one-dimensional site index labelling sites along
the longitudinal direction of the applied potential gradient (the
transverse degrees of freedom are treated in a mean-field
approximation and so there is no dependence on the transverse site
label), $p$ and $h$ are quasiparticle and quasihole annihilation
operators, $Z$ is the number of nearest neighbors in the
transverse directions and $\mu_\ell$ denotes chemical potential
which enforces the constraints
\begin{equation} \langle
p_{\ell+1}^{\dagger} p_{\ell+1}\rangle  = \langle h_\ell^{\dagger}
h_\ell\rangle. \label{const1}
\end{equation}
Although the Hamiltonian (\ref{ham3d}) has no non-linear terms,
its diagonalization is non-trivial because of the hard-core
constraint on all sites
\begin{equation}p_\ell^{\dagger}p_\ell
\le 1,~~~~h_\ell^{\dagger} h_\ell \le 1,~~~~p_\ell^{\dagger}
p_\ell h_\ell^{\dagger} h_\ell =0.
\label{const2}
\end{equation}
The mean fields $\langle p_\ell\rangle $ and $\langle
h_\ell\rangle $ correspond to transverse particle/hole superfluid
order and were self consistently determined by diagonalizing the
3D Hamiltonian (\ref{ham3d}) while maintaining (\ref{const2}).

We now consider the evolution of the ground state under a sudden
shift in the value of $E$ from $E=E_i$ to $E=E_f$ at $t=0^+$. We
place $E_i$ in a regime where the ground state preserves all
symmetry, and there is neither Ising or transverse superfluid
order. The initial ground state $|\Psi^{\rm 3D}\rangle $ will
evolve according to the new Hamiltonian $H_{\rm 3D}[\langle
p\rangle ,\langle h\rangle ;E_f]$. However, in contrast to the 1D
case, here the evolutions of the mean fields $\langle p\rangle $
and $\langle h\rangle $ have to be self-consistently determined.
Within time-dependent Hartree approximation, we obtain
\begin{eqnarray}
|\Psi^{\rm 3D}(t)\rangle  &=& \sum_m c_m(t) |m\rangle  \nonumber\\
i \hbar \frac{d c_m (t)}{dt} &=&\sum_n c_n(t) \nonumber\\
&& \times \langle n|H_{\rm 3D}[\langle p_\ell(t)\rangle ,\langle h_\ell(t)\rangle ;E_f]|m\rangle  \nonumber\\
\langle p_\ell(t)\rangle  &=& \sum_{m,n} c_m^*(t) c_n(t) \langle m|p|n\rangle  \nonumber\\
\langle h_\ell(t)\rangle  &=& \sum_{m,n} c_m^*(t) c_n(t) \langle
m|h|n\rangle .
 \label{scd}
\end{eqnarray}
We used a basis of states $|n\rangle $ (the final results are, of
course, independent of the choice of this basis) which are the
complete set of eigenkets of the Hamiltonian $H_{\rm 3D}[\langle
p_\ell^f\rangle ,\langle h_\ell^f\rangle ;E_f]$, where $\langle
p_\ell^f\rangle $ and $\langle h_\ell^f\rangle $ are the ground
states values of the particle and hole order condensates for
$E=E_f$. All the states $|n \rangle$ maintain (\ref{const2})
exactly, and so these hard-core constraints are fully respected by
our calculation: this is what makes diagonalization of the
Hamiltonian time consuming and numerically intensive. We note that
these equations also maintain the constraints (\ref{const1}) at
all $\ell$ and $t$.

We examined the above equations for the transverse superfluid
order using the same protocol used in Section~\ref{oned} for the
Ising order. The set of Eqs.\ \ref{scd} were solved
self-consistently for longitudinal system size $N=4$. We consider
the starting potential gradient $E_i$ to be in the insulator phase
with neither superfluid or Ising order, and ramp up the potential
gradient to enter the superfluid phase. The gauge symmetry of the
superfluid order parameter is broken by adding a small symmetry
breaking term $H_{\rm sym} = -\sum_\ell \eta [(p_\ell + h_\ell) +
{\rm h.c}]$, where $\eta$ is an infinitesimal positive constant.
In the absence of such a symmetry breaking term we would have
$\langle p_{\ell} \rangle$ and $\langle h_{\ell} \rangle$ vanish
identically for all $t$ by gauge invariance. However, even an
infinitesimal symmetry breaking is sufficient, in the suitable
parameter regime, to induce appreciable values of the superfluid
order. In practice, this symmetry breaking is provided by the
coupling of the system to its environment.

\begin{figure}
        \centerline{\epsfig{file=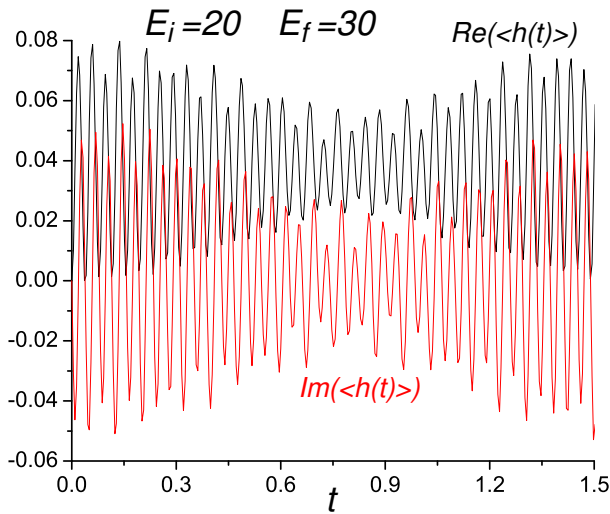,width=8cm,angle=0}}
        \caption{Oscillations of the hole superfluid order parameter.
        The plot is shown for $U=40$, $w=1$, $n_0 = 1$, $Z=4$, $E_i=20$ and $E_f=30$. For
        these parameters, the quantum critical point is at $E_c = 26.4$.}
        \label{fig4a}
\end{figure}

With the symmetry breaking term present, the superfluid order
parameters $\langle p_\ell \rangle$ and $\langle h_\ell \rangle$
are initially real. As seen in Fig.\ \ref{fig4a}, $\langle
h(t)\rangle = \sum_\ell \langle h_\ell(t)\rangle $ develop
coherent oscillations once the potential gradient takes the value
$E_f$. (These results are the analog of
Figs~\ref{fig1},\ref{fig2}.)
The oscillations in $\langle p(t)\rangle $ are similar, but occur
with a different period due to the inherent particle-hole
asymmetry in Eq.\ \ref{ham3d}. The long time average of the
oscillations is purely real, and this is, of course determined by
the symmetry breaking term. Such oscillations of the superfluid
order parameter were also obtained recently in a different context
in Refs.~\onlinecite{aa,psg} and by Levitov\ \cite{Levitov}.

\begin{figure}
        \centerline{\epsfig{file=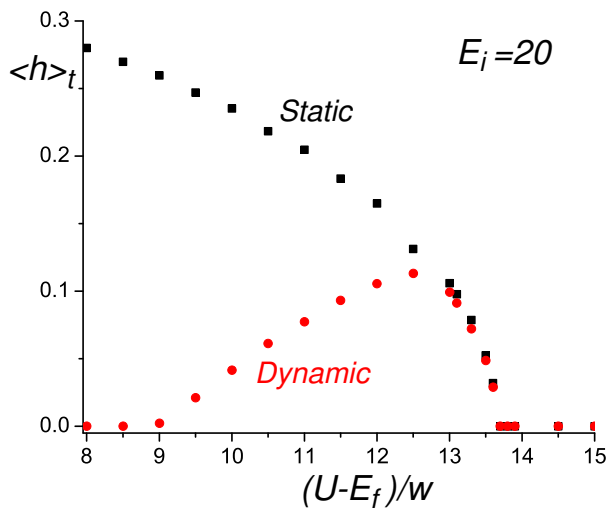,width=8cm,angle=0}}
        \caption{Analog of Fig~\protect\ref{fig3}, but for the
        transverse superfluid order, using the same parameters
        (apart from $E_f$) as Fig~\protect\ref{fig3a}. The
        "static" curve is the equilibrium superfluid order
        parameter determined in I. The "dynamic" curve is the
        long time average of the real part of $\langle h(t)
        \rangle$.}
        \label{fig4b}
\end{figure}

We also examined the $E_f$ dependence of the long time average of
the superfluid order, and the analog of Fig~\ref{fig3} appears in
Fig~\ref{fig4b}.
Again the superfluid order is most strongly enhanced in the
vicinity of the quantum critical point. However, unlike
Fig~\ref{fig3}, we do not observe a precursor to the superfluid
order in the insulating phase: this is surely an artifact of the
mean-field treatment of the transverse degrees of freedom.

\section{Conclusions}
\label{conc}

With advent of the study of quantum phase transitions in trapped
atomic systems, there is a clear need for theoretical studies in
the highly non-equilibrium situations that experiments are often
in. In particular, experiments can easily explore the change in
the state of the system upon a sudden change in a parameter in the
Hamiltonian. There are few general principles in such cases ({\em
e.g.\/} there is no fluctuation-dissipation theorem which controls
correlations of the final state), and theory is clearly still in
its infancy. Two recent studies in this class \cite{aa,psg},
examined the evolution of superfluid order under a sudden change
in the optical lattice potential exerted on trapped bosons.

It is clear that exact results on simple solvable models in
non-equilibrium situations would be valuable. We have provided
such an example here in Section~\ref{sec:ising}, where we examined
the Ising chain in a transverse field, $g$. This model has a
quantum critical point at $g=g_c$, with spontaneous ferromagnetic
order in the ground state for $g<g_c$. We started the Ising model
in the paramagnetic state ($g_i \gg g_c$), suddenly at $t=0$
changed $g$ to a final value $g_f$, and examined the long time
development of correlations of the ferromagnetic order. (Our
formalism also provided results for all $t>0$, but we have not
examined the detailed time evolution here.) The results are
summarized in Fig~\ref{fig3b}. True long-range order does not
develop at any value of $g_f$; however, significant order
parameter correlations do appear, and these are best formed for
$g_f \approx g_c$. In a general non-integrable system we may
expect thermalization at long times, at a temperature such that
the average energy equals that of the state at $t=0^+$. Such
thermalization does not occur for the present integrable system,
and the results have certain artifacts associated with this: the
long-time correlations have an oscillatory spatial dependence for
$g_f < g_c$.

In the remainder of the paper we studied the non-equilibrium
dynamics of models introduced in a previous paper \cite{Sachdev2}
which addressed the response of a bosonic Mott insulator to a
change in a strong potential gradient \cite{Bloch}. These models
exhibit a number of quantum critical points associated with the
onset of Ising density wave and superfluid order. Our numerical
studies here found a feature similar to that also obtained for the
solvable Ising model: the order was best formed when the final
parameter value was in the vicinity of the associated quantum
critical point, as illustrated in Fig~\ref{fig3}. Here, and in
Ref.~\onlinecite{Sachdev2}, we have proposed this feature as the
explanation for the resonant response observed by Greiner {\em et
al.} \cite{Bloch} upon `tilting' a Mott insulator of bosons in an
optical lattice.

\begin{acknowledgments}
We thank A. G. Abanov, S.~M.~Girvin, L.~Levitov, and
A.~Polkovnikov for useful discussions. This research was supported
by US NSF grants DMR-0098226 and DMR-0342157.
\end{acknowledgments}

\appendix*

\section{Computations for the Ising chain}

The Jordan-Wigner transformation allows the Hamiltonian of an
Ising chain in a transverse field $g$ to be written as \be
\label{Hamiltonian} \Ham_I = \sum_k \epsilon_k \gamma_k^\dag
\gamma_k, \ee where $\gamma_k$ is a fermionic annihilation
operator (see Chapter 4 of Ref.~\onlinecite{Sachdev1}). These are
related to the Jordan-Wigner fermions $c_k$ by a Bogoliubov
transformation, parametrized by angle $\theta_k$, where
\be\label{DefineThetaK} \tan\theta_k = \frac{\sin k}{g - \cos k}.
\ee In the present case, we define the $\gamma$ fermions as those
that diagonalize the Hamiltonian for $t > 0$, with field $g_f$.

Since the Hamiltonian is throughout translationally symmetric,
only fermionic states with opposite pseudomomentum $k$ and $-k$
are mixed. We may therefore write the two component column vector
\be\label{DefineGammaK}\Gamma_k =
\begin{pmatrix}
\gamma_k\\
\gamma_{-k}^\dag
\end{pmatrix}
\ee and similarly $c_k$ for the Jordan-Wigner fermions. The
Bogoliubov transformation relating $c_k$ and $\Gamma_k$ is
expressed as $c_k = R^x(\theta_k) \Gamma_k$, where \be R^x(\alpha)
= \cos\frac{\alpha}{2} + i\sigma^x\sin\frac{\alpha}{2} \ee and
here $\sigma^x$ is a $2\times 2$ Pauli matrix. (These are used for
conciseness of notation and should not be confused with the
operators representing the `spins' of the Ising chain.)

For $t < 0$, the field is $g_i$ and the system is taken to be in
its ground state. We define the $\gamma'$ fermions as those which
diagonalize the Hamiltonian in the form (\ref{Hamiltonian}) with
this field. (Similarly, $\theta_k'$ and $\Gamma_k'$ are given by
analogy with (\ref{DefineThetaK}) and (\ref{DefineGammaK}).)

The state $|\psi\rangle$ is therefore the vacuum of $\gamma'$
fermions: in matrix notation, \be \langle\psi |
\Gamma_k'\Gamma_k'^\dag|\psi\rangle =
\begin{pmatrix}
1 & 0\\
0 & 0
\end{pmatrix}
= \frac{1}{2}(\sigma^z + 1). \ee Applying the Bogoliubov
transformation to $c_k$ and then $\Gamma_k$ gives \be \langle\psi
| \Gamma_k\Gamma_k^\dag|\psi\rangle = R^{x\dag}(\theta_k -
\theta_k') \frac{1}{2}(\sigma^z + 1) R^x(\theta_k - \theta_k').
\ee Using $R^{x\dag}(\alpha)\sigma^z R^x(\alpha) = \sigma^z \cos
\alpha - \sigma^y \sin \alpha$ with $\phi_k = \theta_k -
\theta_k'$ gives \be \langle\psi |
\Gamma_k\Gamma_k^\dag|\psi\rangle = \frac{1}{2}(1 + \sigma^z \cos
\phi_k - \sigma^y \sin \phi_k), \ee as the set of matrix elements
for the initial state.

The time evolution of the operators now proceeds (using the
Heisenberg picture) according to the Hamiltonian,
(\ref{Hamiltonian}), so that $\Gamma_k(t) = U_k(t) \Gamma_k(0)$,
where \be U_k(t) =
\begin{pmatrix}
e^{-i \epsilon_k t} & 0\\
0 & e^{i \epsilon_k t}
\end{pmatrix}
= R^{z\dag}(2\epsilon_k t). \ee The expectation values at any time
can therefore be evaluated using the algebra of $SU(2)$ matrices.

The $n$-site correlator can be written as $\langle G_n \rangle$ =
$\langle B_0 A_1 B_1 \ldots B_{n-1} A_n \rangle$, where $A_i =
c_i^\dag + c_i$ and $B_i = c_i^\dag - c_i$. Wick's theorem can
then be used to write this expression in terms of the expectation
values of expressions bilinear in $A_i$ and $B_i$. We therefore
let \be \Omega_k(t) =
\begin{pmatrix}
A_k(t)\\
B_k(t)
\end{pmatrix}
= \sqrt{2} R^{y}\left(\frac{\pi}{2}\right) C_k(t), \ee where $A_k$
($B_k$) is the Fourier transform of $A_i$ ($B_i$) so that
\begin{widetext}
\bea \langle\psi | \Omega_k \Omega_k^\dag|\psi\rangle &=&
\langle\psi |
\begin{pmatrix}
A_k A_{-k} & -A_k B_{-k} \\
B_k A_{-k} & -B_k B_{-k}
\end{pmatrix}
|\psi\rangle = 2 R^{y}\left(\frac{\pi}{2}\right)\langle\psi | C_k C_k^\dag|\psi\rangle R^{y\dag}\left(\frac{\pi}{2}\right)\\
&=& 1 - \sigma^x (\cos \theta_k \cos \phi_k - \sin\phi_k\sin\theta_k\cos 2\epsilon_k t)\nonumber\\
& &+ \sigma^y (\sin \theta_k \cos \phi_k +
\sin\phi_k\cos\theta_k\cos 2\epsilon_k t)
+ \sigma^z (\sin\phi_k \sin 2\epsilon_k t)\\
&=& \begin{pmatrix}
1 - \sin\phi_k\sin 2\epsilon_k t & -e^{i \theta_k} (\cos \phi_k + i \sin\phi_k\cos 2\epsilon_k t)\\
-e^{-i \theta_k} (\cos \phi_k - i \sin\phi_k \cos 2\epsilon_k t) &
1 + \sin\phi_k\sin 2\epsilon_k t
\end{pmatrix}.
\eea
\end{widetext}

Transforming to real space and using the conservation of
pseudomomentum gives \bea
\langle A_lA_j\rangle &=& \frac{1}{M}\sum_k e^{ik(l-j)}(1-\sin\phi_k\sin 2\epsilon_k t)\nonumber\\
\langle B_lB_j\rangle &=& \frac{1}{M}\sum_k e^{ik(l-j)}(-1-\sin\phi_k\sin 2\epsilon_k t)\nonumber\\
\langle B_lA_j\rangle &=& \frac{1}{M}\sum_k e^{ik(l-j)}e^{-i\theta_k}(-\cos\phi_k\nonumber\\
& &+ i\sin\phi_k\cos 2\epsilon_k t). \eea The long time averages
of these expressions are \bea
\langle A_l A_j \rangle &=& \delta_{lj}\nonumber\\
\langle B_l B_j \rangle &=& -\delta_{lj}\nonumber\\
\langle B_l A_j \rangle &=& a_{l-j+1}, \eea where \bea
a_r &=& \frac{1}{M}\sum e^{-ikr} \tilde{a}(k)\\
&=& \frac{1}{2\pi}\int_{-\pi}^{\pi} e^{-ikr} \tilde{a}(k), \eea in
the limit where the number of sites $M$ becomes infinite. Here,
$\tilde{a}(k) = -e^{i(\theta_k + k)}\cos (\theta_k - \theta_k')$.
Wick's theorem then allows the expression for $\langle G_n
\rangle$ to be written as (\ref{toep}).

\end{document}